\documentclass[aps,pre,reprint,twocolumn,showpacs,superscriptaddress,groupaddress,showkeys,floatfix,byrevtex]{revtex4-1}
\usepackage{amsmath}
\usepackage{amsfonts}
\usepackage{amssymb}
\usepackage[dvipsnames]{xcolor}
\usepackage{graphicx,color,subeqnarray}
\usepackage{dcolumn}
\usepackage{bm}
\usepackage{soul}

\usepackage{srcltx}
\makeatletter
\DeclareFontFamily{OMX}{MnSymbolE}{}
\DeclareSymbolFont{MnLargeSymbols}{OMX}{MnSymbolE}{m}{n}
\SetSymbolFont{MnLargeSymbols}{bold}{OMX}{MnSymbolE}{b}{n}
\DeclareFontShape{OMX}{MnSymbolE}{m}{n}{
    <-6>  MnSymbolE5
   <6-7>  MnSymbolE6
   <7-8>  MnSymbolE7
   <8-9>  MnSymbolE8
   <9-10> MnSymbolE9
  <10-12> MnSymbolE10
  <12->   MnSymbolE12
}{}
\DeclareFontShape{OMX}{MnSymbolE}{b}{n}{
    <-6>  MnSymbolE-Bold5
   <6-7>  MnSymbolE-Bold6
   <7-8>  MnSymbolE-Bold7
   <8-9>  MnSymbolE-Bold8
   <9-10> MnSymbolE-Bold9
  <10-12> MnSymbolE-Bold10
  <12->   MnSymbolE-Bold12
}{}

\let\llangle\@undefined
\let\rrangle\@undefined
\DeclareMathDelimiter{\llangle}{\mathopen}
{MnLargeSymbols}{'164}{MnLargeSymbols}{'164}
\DeclareMathDelimiter{\rrangle}{\mathclose} {MnLargeSymbols}{'171}{MnLargeSymbols}{'171}
\makeatother

\begin{document}
\title{Generalized Ornstein-Uhlenbeck Model for Active Motion}
\author{Francisco J. \surname{Sevilla}}
\email{fjsevilla@fisica.unam.mx}
\thanks{corresponding author}
\affiliation{Departamento de Sistemas Complejos,
Instituto de F\'isica,
Universidad Nacional Aut\'onoma de M\'exico,\\
Apdo. Postal 20-364, 01000, Ciudad de M\'exico, M\'exico}

\author{Rosal\'io F. \surname{Rodr\'iguez}}
\affiliation{Departamento de Sistemas Complejos, Instituto de F\'isica, Universidad Nacional Aut\'onoma de M\'exico,\\
Apdo. Postal 20-364, 01000, Ciudad de M\'exico, M\'exico}
\affiliation{FENOMEC, Universidad Nacional Aut\'onoma de M\'exico, Apdo. Postal 20-726, 01000, Ciudad de M\'exico, M\'exico}

\author{Juan Ruben \surname{Gomez-Solano}}
\affiliation{Departamento de Sistemas Complejos, Instituto de F\'isica, Universidad Nacional Aut\'onoma de M\'exico,\\
Apdo. Postal 20-364, 01000, Ciudad de M\'exico, M\'exico}

\begin{abstract}
We investigate a one-dimensional model of active motion, which takes into account the effects of persistent self-propulsion through a memory function in a dissipative-like 
term of the generalized Langevin equation for particle swimming velocity. The proposed model is a generalization of the active Ornstein-Uhlenbeck model introduced by G. 
Szamel [Phys. Rev. E {\bf 90}, 012111 (2014)]. We focus on two different kinds of memory which arise in many natural systems: an exponential decay and a power law, 
supplemented with additive colored noise. We provide analytical expressions for the velocity autocorrelation function and the mean-squared displacement, which are in 
excellent agreement with numerical simulations. For both models, damped oscillatory solutions emerge due to the competition between the memory of the system and the 
persistence of velocity fluctuations. In particular, for a power-law model with fractional Brownian noise, we show that long-time active subdiffusion occurs with increasing 
long-term memory. 
\end{abstract}

\date{\today}

\maketitle

\section{Introduction}

Systems in out-of-equilibrium conditions are ubiquitous in nature, among which biological
active matter is the most representative. For instance, motile bacteria employ diverse 
swimming patterns to traverse complex 
habitats \cite{TaktikosPlos2013,TauteNatComm2015}. Recent technological advances have allowed the design of artificial particles that take 
advantage of different physical and/or chemical mechanisms to self-induce motion that mimics biological motility \cite{BechingerRMP2016}. Such mobile entities, 
either biological \cite{RamaswamyAnnRevCondMattPhys2010,MarchettiRMP2013} or human-made \cite{HowsePRL2007,PalacciPRL2010,JiangPRL2010,GaoACSNAno2015,GomezSolanoSciRep2017}, 
are able to
develop autonomously directed motion by using the locally available energy from the environment \cite{BechingerRMP2016}. These particles are called \emph{self-propelled} or 
more generally, \emph{active particles}. 

For nonequilibrium statistical physicists, active matter provides a rich field of research that has allowed the rapid progress of different theoretical frameworks. It has 
been pointed out that the detailed balance between the injection and the dissipation of energy is not satisfied at the 
microscopic scale in active systems. However, many of the accomplished advancements in the understanding of active matter have partly relied on the intuition built from 
equilibrium systems \cite{TakatoriPRL2014,GinotPRX2015,TakatoriPRE2015}. For instance, the concept of \emph{effective temperature} has provided a valuable description 
of some out-of-equilibrium systems \cite{OukrisNphys2010, ColombaniPRL2011,DieterichNaturePhys2015}, and in particular in systems of active particles 
\cite{LoiPRE2008,TailleurEPL2009,PalacciPRL2010,EnculescuPRL2011,BenIsaacPRL2011,LoiSoftMatterM2011,SzamelPRE2014,LevisEPL2015,SevillaPRE2019}. In general, the possibility of defining 
 an effective temperature relies on the fulfillment of a nonthermal fluctuation-dissipation relation. This is the case for timescales larger than the persistence one, for 
which the motion of free active particles is well characterized by an effective diffusion coefficient. Such a behavior can be interpreted as the motion of a passive Brownian 
particle diffusing in a \emph{fictitious} environment at an effective temperature higher than the true equilibrium temperature of the surroundings.

A model of active motion that has attracted a great deal of attention because of its simplicity is the so-called \emph{active Ornstein-Uhlenbeck model} (AOUM). It is based on 
the assumptions that in the overdamped regime, the particle position changes in time due to all the potentials that affect its motion, as well as due to 
its own self-propulsion velocity, which is described by an Ornstein-Uhlenbeck process \cite{SzamelPRE2014}. The AOUM has been used as a basis to consider interactions among 
self-propelled particles \cite{SzamelPRE2015,MarconiSoftMatter2015} and to study the main nonequilibrium features exhibited by active matter, such as \emph{motility-induced 
phase 
separation} \cite{FaragePRE2015,FodorPRL2016}. Also, it has allowed the derivation of analytical results in the case of independent active particles confined in simple 
potentials \cite{DasNJP2018,CapriniJStatMech2018}. Furthermore, within the framework of stochastic thermodynamics, it has permitted the analysis  of entropy production, 
fluctuation theorems, and Clausius relations for active matter \cite{MandalPRL2017,PuglisiEntropy2017,MarconiSciRep2017}.

In this paper we consider a generalization of the AOUM based on the \emph{generalized Langevin equation} (GLE) \cite{KuboRepProgPhys1966,FoxJMathPhys1977}, which endows the 
standard Langevin model of Brownian motion with finite time correlations. The GLE usually models systems in viscoelastic baths near equilibrium states and includes retarded 
memory effects in the viscous drag term of the equation and correlated thermal 
noises~\cite{WangPhysA1999,Pottier2003,ViñalesPRE2007,DespositoPRE2008,DespositoPRE2009,FigueiredoJMathPhys2009,SandevPhysScr2010,SandevJMathPhys2014}. 
Remarkably, these kinds of models are also of great theoretical interest to describe nonequilibrium systems, as memory effects cannot be neglected in many situations. For 
active matter, memory effects can significantly alter the directional dynamics of individual self-propelled particles when moving in viscoelastic media. For instance, in 
polymer solutions
the persistence length of flagellated bacteria \cite{PattesonSciRep2015} and synthetic nanopropellers \cite{SchamelACSNano2014} is enhanced, while self-propelled spherical 
colloids exhibit an increase of rotational diffusion \cite{GomezSolanoPRL2016} and circular trajectories \cite{NarinderPRL2018}. Memory effects are revealed in many other 
active systems with long-range temporal correlations that also motivate our analysis, e.g., self-propelled particles in glassy~\cite{HenkesPRE2011} or disordered 
heterogeneous media~\cite{ChepizhkoPRL2013,MorinPRE2017}, motile bacteria with intricate swimming patterns~\cite{TaktikosPlos2013}, microorganisms with strong autochemotactic 
response~\cite{TaktikosPRE2011}, and active liquid-crystal droplets~\cite{SugaPRE2018}.

In Sec. \ref{SectII} we present the explicit formulation of the model that describes the motion of self-propelled particles subject to thermal and active fluctuations. We show 
that the probability density of the complete process can be written as the convolution of the diffusion probability density, due to thermal fluctuations, and the corresponding 
probability distribution of the active part of motion, which is analyzed in Sec. \ref{SectIII}. In the same section two relevant examples are discussed in detail, first, a 
memory function that models the  retarded  effects  on  the  swimming  velocity  due to  viscoelastic-like  effects, and, second, a memory function with power-law long-lived 
correlations. Both examples qualitatively capture the phenomenology observed in a variety of active systems, namely the occurrence of anticorrelations of 
the swimming velocity which lead to self-trapping effects. Finally, in Sec. \ref{SectIV} we summarize the main results of our work and make some further physical remarks. 

\section{\label{SectII}The Generalized Ornstein-Uhlenbeck Model of Active Motion}
One remarkable aspect of the motion of active particles is that it is \emph{persistent}, i.e., the particles approximately retain the state of motion for a characteristic 
finite timescale, called the \emph{persistence time}. This feature is indeed observed in the patterns of motion of different microorganisms and some artificially designed 
self-motile particles. For instance, the \emph{run-and-tumble} pattern of \emph{Escherichia coli} alternates time intervals at a rather constant speed in a straight line 
along a randomly chosen direction, interrupted by short time periods during which the bacterium tumbles almost at rest. On a statistical description, the run-and-tumble motion 
can be characterized by a finite timescale of persistence, which makes the motility behavior strongly correlated in time, thus rendering the nonequilibrium signatures 
conspicuously observable. 

Here we provide a theoretical framework with the possibility of considering a variety of patterns of  persistent motion. The equations that describe the time evolution of the particle position $x(t)$ of an overdamped active Brownian particle diffusing in one dimension, and the time evolution of its swimming velocity, $v_{\text{s}}(t)$, are 
given by
\begin{subequations}\label{gle}
\begin{align}
\frac{d}{dt}x(t)&=v_{\text{s}}(t)+\xi_{x}(t),\label{glex}\\
\frac{d}{dt}v_{\text{s}}(t)&=- \frac{1}{\tau_R}\int_{0}^{t}ds\, \gamma(t-s)v_{\text{s}}(s)+\xi_{v_{\text{s}}}(t).\label{glev}
\end{align}
\end{subequations}
In Eq. \eqref{glex}, $\xi_{x}(t)$ denotes the thermal noise caused by the medium, which is modeled here as Gaussian white noise, i.e., with average 
 $\langle\xi_{x}(t)\rangle=0$ and autocorrelation function $\langle\xi_{x}(t)\xi_{x}(s)\rangle=2D_{T}\delta(t-s)$; $D_{T}$ is the diffusion constant due to translational 
motion given by $\mu k_{B}T$, $\mu$ being the mobility; $k_{B}$ the Boltzmann constant; and $T$ the medium temperature. Equation \eqref{glev} is the well-known 
GLE that in the context of the present paper provides a generalization of the AOUM of active motion \cite{SzamelPRE2014}, which takes into account the exponential 
correlations of the swimming velocity that gives rise to exponentially persistent motion. Here, Eq. \eqref{glev} opens the door for taking into account a variety of 
persistent motions by properly choosing the memory function $\gamma(t)$ \cite{SevillaChapter2018}, which has units of [time]$^{-1}$. The timescale $\tau_R$ in 
Eq.~(\ref{glev}) characterizes the persistence of the velocity fluctuations (the \emph{persistence time}). For times larger than $\tau_R$, they relax to zero, fading out the 
ballistic motion. 

We focus on the physically relevant case where Eqs. \eqref{gle} describe a stationary process whose statistical properties are invariant under temporal 
translations. For simplicity, the \emph{noise} term $\xi_{v_{\text{s}}}(t)$ is assumed to be stationary and Gaussian with vanishing 
average $\langle\xi_{v_{\text{s}}}(t)\rangle=0$ and autocorrelation function 
\begin{equation}\label{NoiseACF}
 \langle\xi_{v_{\text{s}}}(t)\xi_{v_{\text{s}}}(s)\rangle=\frac{v_0^2}{\tau_R}\eta(\vert t-s\vert).
\end{equation}
In Eq.~(\ref{NoiseACF}), $\eta(t)$ is a function with physical units of time$^{-1}$, whereas $v_0$ determines the variance of the velocity fluctuations, $\langle v_{\text{s}}(t) 
v_{\text{s}}(t) \rangle = \langle v_{\text{s}}(0) v_{\text{s}}(0) \rangle = v_0^2$, which defines the \emph{characteristic self-propelling speed} $v_{0}$. Although there are 
no \emph{a priori} reasons {to establish a} relation between $\gamma(t)$ and $\eta(t)$, it is physically plausible that the relation 
{$\eta(t)=\gamma(t)$} may be sustained in some cases of interest. {This relation does not imply thermal equilibrium but only expresses the 
simple situation, described by linear-response theory, for which the response of the swimming velocity to active fluctuations is connected by the square of the 
self-propelling speed divided by the persistent time }\cite{Note1}. The active Ornstein-Uhlenbeck  model of Szamel \cite{SzamelPRE2014} is recovered {from Eq. 
\eqref{glev}} for the zero-ranged memory function $\gamma(t)=\eta(t)=  2 \delta(t)$, which leads to an exponentially decaying autocorrelation function, i.e., $\langle v_s(t) 
v_s(s) \rangle = v_0^2 \exp(-|t - s|/\tau_R)$, also considered in the analysis of a two-dimensional active motion in Ref. \cite{GhoshJChemPhys2015}.

We pay particular attention to the statistical properties of active motion induced by finite- and long-ranged memory functions. We are mainly interested on the statistics of the particle swimming velocity an its position, for which the explicit dynamics 
of the self-propulsion velocity is implied by the memory function $\gamma(t)$. The formal solutions of Eqs. \eqref{gle} are given explicitly by 
\begin{subequations}\label{glesols}
\begin{align}
x(t)&=\llangle x(t)\rrangle +\int_{0}^{t}ds\, \Gamma(t-s)\xi_{v_{\text{s}}}(s)+\int_{0}^{t}ds\, \xi_{x}(s),\label{glexsol}\\
v_{\text{s}}(t)&=\langle v_{\text{s}}(t)\rangle+\int_{0}^{t}ds\, \Gamma^{\prime}(t-s)\xi_{v_{\text{s}}}(s)\label{glevsol},
\end{align}
\end{subequations}
where 
\begin{subequations}\label{AverageTrajectories}
\begin{align}
    \llangle x(t)\rrangle&=x(0)+v_{\text{s}}(0)\,\Gamma(t),\label{Gamma}\\
    \langle v_{\text{s}}(t)\rangle&=v_{\text{s}}(0)\,\Gamma^{\prime}(t),\label{GammaPrime}
\end{align}
\end{subequations}
give the mean position and the mean swimming velocity, respectively. The average $\llangle\cdot\rrangle$ is taken over the independent 
realizations of the Gaussian white noises $\xi_{x}(t)$ and $\xi_{v_{\text{s}}}(t)$, while $\langle\cdot\rangle$ only over realizations of $\xi_{v_{\text{s}}}(t)$. 
$x(0)$ and $v_{\text{s}}(0)$ are the corresponding initial values. $\Gamma(t)$ and $\Gamma^{\prime}(t)=d\Gamma(t)/dt$ are the solutions of the deterministic counterpart of Eqs. \eqref{glex} and \eqref{glev} and given by the inverse Laplace transform of  
\begin{subequations}\label{Gammas}
\begin{align}
\widetilde{\Gamma}(\epsilon)&=\epsilon^{-1}\widetilde{\Gamma^{\prime}}(\epsilon)
,\label{LTGamma}\\
\widetilde{\Gamma^{\prime}}(\epsilon)&=\left[\epsilon+\frac{1}{\tau_{R}}\widetilde{\gamma}(\epsilon)\right]^{-1},\label{LTGammaPrime}
\end{align}
\end{subequations}
respectively. The symbol $\widetilde{f}(\epsilon)$ denotes the Laplace transform of the function of time $f(t)$, defined by 
$\widetilde{f}(\epsilon)=\int_{0}^{\infty}dt\, e^{-\epsilon t}f(t)$ with $\epsilon$ the Laplace variable, a complex number.

The long-time regime of the quantities \eqref{Gammas} is determined by the asymptotic behavior of $\gamma(t)$. It is customary to require that $\gamma(t)$ vanishes with increasing $t$, which means that in the Laplace domain 
$\lim_{\epsilon\rightarrow0}\epsilon\widetilde{\gamma}(\epsilon)\rightarrow 0$. A necessary and sufficient condition for a well-defined asymptotic limit of $\Gamma(t)$ and 
$\Gamma^{\prime}(t)${, and therefore a well-behaved time dependence of the average trajectories \eqref{AverageTrajectories}}, is that $\epsilon\widetilde{\gamma}(\epsilon)$ 
goes to zero slower than $\epsilon^{2}$. This is trivially satisfied by positive monotonically decreasing memory functions{---which maintain 
the physical interpretation of persistence---}that go exponentially or faster to zero or by those that go to zero as $t^{-\beta}$ with $0<\beta<1$. 

The characteristic function of the probability density associated to the stochastic process defined by Eqs. \eqref{gle} is given by
\begin{multline}
    \hat{G}(k,q,t)=\left\llangle \exp\left\{-i\int_{0}^{t}ds\, k\,  \xi_{x}(s)\right\}\times\right.\\
    \left. \exp\left\{-i\int_{0}^{t}ds\,\Bigl[q\Gamma^{\prime}(t-s)+k\Gamma(t-s)\Bigr]\xi_{v_{\text{s}}}(s)\right\}\right\rrangle.
\end{multline}
This quantity can be explicitly written as the product of the characteristic function of the translational part $\hat{G}_{D_{T}}(k,t)$ times the corresponding bivariate characteristic 
function of the active part $\hat{G}^{(2)}_{\text{act}}(k,q,t)$, i.e.,
\begin{equation}\label{PDFFactorization}
    \hat{G}(k,q,t)=\hat{G}_{D_{T}}(k,t)\, \hat{G}^{(2)}_{\text{act}}(k,q,t),
\end{equation}
where
\begin{subequations}
\begin{align}\label{TransG}
    \hat{G}_{D_{T}}(k,t)=\exp\left\{-D_{T}k^{2}t\right\}
\end{align}
is the univariate characteristic function of the diffusion equation, $D_{T}$ the diffusion coefficient linked to thermal fluctuations, and
\begin{multline}\label{ActG}
    \hat{G}^{(2)}_{\text{act}}(k,q,t)=\exp\left\{-\frac{1}{2}q^{2}\sigma_{v_{\text{s}}v_{\text{s}}}^{2}(t)\right.\\ 
\left.-qk\sigma_{xv_{\text{s}}}^{2}(t)
-\frac{1}{2}k^{2}\sigma_{xx}^{2}(t)\right\},
\end{multline}
\end{subequations}
is a bivariate Gaussian that corresponds to the characteristic function of active motion. 
The expression for $\hat{G}^{(2)}_{\text{act}}(k,q,t)$ in Eq.~(\ref{ActG}) explicitly involves the standard elements of the active covariance matrix $\Sigma_{\text{act}}$, i.e., the variance of the particle position $\sigma^{2}_{xx}(t)\equiv\left\langle\left[x(t)-\langle x(t)\rangle\right]^{2}\right\rangle$, the variance of the particle swimming velocity 
$\sigma^{2}_{v_{\text{s}}v_{\text{s}}}(t)\equiv\left\langle\left[v_{\text{s}}(t)-\langle v_{\text{s}}(t)\rangle\right]^{2}\right\rangle$, and the covariance of the particle position and swimming velocity $\sigma^{2}_{xv_{\text{s}}}\equiv\Bigl\langle\left[x(t)-\langle x(t)\rangle\right] \left[v_{\text{s}}(t)-\langle 
v_{\text{s}}(t)\rangle\right]\Bigr\rangle$. Such matrix elements are given by
\begin{subequations}\label{CovMatrixElements}
\begin{align}
\sigma_{v_{\text{s}}v_{\text{s}}}^{2}(t)&=\frac{v_{0}^{2}}{\tau_{R}}\int_{0}^{t}ds_{1}\int_{0}^{t}ds_{2}\, \Gamma^{\prime}(s_{1})\, \Gamma^{\prime}(s_{2})\, \eta(\vert 
s_{1}-s_{2}\vert),\\
\sigma_{xx}^{2}(t)&=\frac{v_{0}^{2}}{\tau_{R}}\int_{0}^{t}ds_{1}\int_{0}^{t}ds_{2}\, \Gamma(s_{1})\, \Gamma(s_{2})\, \eta(\vert 
s_{1}-s_{2}\vert),\\
\sigma_{xv_{\text{s}}}^{2}(t)&=\frac{v_{0}^{2}}{\tau_{R}}\int_{0}^{t}ds_{1}\int_{0}^{t}ds_{2}\, \Gamma^{\prime}(s_{1})\, \Gamma(s_{2})\, \eta(\vert 
s_{1}-s_{2}\vert),
\end{align}
\end{subequations}
and are valid for arbitrary $\gamma(t)$ and $\eta(t)$.

Thus, the joint probability density of finding a particle at position $x$ and swimming with velocity $v_{\text{s}}$ at time $t$, given that initially  ($t=0$) the particle was located at $x(0)$ swimming at velocity $v_{\text{s}}(0)$, $ P\Bigl(x,v_{\text{s}},t\vert x(0),v_{\text{s}}(0)\Bigr)$,
can be written as the convolution
\begin{multline}\label{PDFsConcolution}
P\Bigl(x,v_{\text{s}},t\vert x(0),v_{\text{s}}(0)\Bigr)=\int_{-\infty}^{\infty}dx^{\prime} G_{D_{T}}\bigl(x-\llangle x(t)\rrangle-x^{\prime},t\bigr)\times\\  
G^{(2)}_{\text{act}}\bigl(x^{\prime},v_{\text{s}}-\langle v_{\text{s}}(t)\rangle,t\bigr),
\end{multline}
where 
\begin{equation}
 G_{D_{T}}\left(x,t\right)=\frac{1}{\sqrt{4\pi D_{T}t}}\exp\left\{-\frac{x^{2}}{4D_{T}t}\right\},
\end{equation}
is obtained straightforwardly by inverting the Fourier transform of Eq. \eqref{TransG},
while 
\begin{widetext}
\begin{equation}
G^{(2)}_{\text{act}}\left(x,v_{\text{s}},t\right)=\frac{1}{2\pi\sigma_{xx}(t)\sigma_{v_{\text{s}}v_{\text{s}}}(t)\sqrt{1-C(t)}}\exp\left\{-\frac{1}{2[1-C(t)]}\left(\frac{v_{
\text { s } } ^ { 2 } } { \sigma_{v_{\text{s}}v_{\text{s}}}^{2}(t)}-\frac{2xv_{\text{s}}\, C(t)}{\sigma_{xv_{\text{s}}}^{2}(t)}+\frac{x^{2}}{\sigma_{xx}^{2}(t)}
\right)\right\}
\end{equation}
\end{widetext}
is obtained after inverting the Fourier transform of \eqref{ActG}, where 
\begin{equation}
 C(t)=\frac{\sigma_{xv_{\text{s}}}^{2}(t)}{\sigma_{v_{\text{s}}v_{\text{s}}}^{2}(t)}\frac{\sigma_{xv_{\text{s}}}^{2}(t)}{\sigma_{xx}^{2}(t)}.
\end{equation}

\section{\label{SectIII}The statistics of the active component of motion}
We have shown that the dynamics is explicitly split into the translational part and the active one [see Eqs. \eqref{PDFFactorization} and \eqref{PDFsConcolution}]. This 
allows us to focus on the statistical properties of the active part of motion. In such a case, it is equivalent to consider Eq. \eqref{gle} with $D_{T}=0$ [$\xi_{x}(t)=0$ for 
all $t$], which reduces to the standard generalized Langevin equation that describes the persistence effects of active motion through the memory function in the dissipative 
term \cite{SevillaChapter2018}. In order to unveil the main consequences of the model proposed, we restrict our analysis to the case of internal noise, i.e., 
$\eta(t)=\gamma(t)$.

In addition to the quantities given in Eqs. \eqref{CovMatrixElements} [evaluated at $\eta(t)=\gamma(t)$], we consider the autocorrelation function of the swimming velocity $\langle v_{\text{s}}(t)v_{\text{s}}(s)\rangle$ which can be written as
\begin{multline}\label{ACFswimmingvel}
\langle v_{\text{s}}(t)v_{\text{s}}(s)\rangle=v_{\text{s}}^{2}(0)\Gamma^{\prime}(t)\Gamma^{\prime}(s)\\ +\frac{v_{0}^{2}}{\tau_{R}}\int_{0}^{t}ds_{1}\int_{0}^{s}ds_{2}\, 
\Gamma^{\prime}(s_{1})\Gamma^{\prime}(s_{2})\gamma(\vert s_{1}-s_{2}\vert),
\end{multline}
for $s\le t$.

The asymptotic behavior of the quantities \eqref{CovMatrixElements} and \eqref{ACFswimmingvel}, is determined by the corresponding one of $\gamma(t)$, which is deduced by requiring a well-behaved time dependence of
$\Gamma(t)$ and $\Gamma^{\prime}(t)$. Such behavior is fulfilled if (a) $\gamma(t)$ vanishes exponentially or faster or if (b) it vanishes as $t^{-\beta}$ with 
$0<\beta<1$. In any case we have that $\sigma_{v_{\text{s}}v_{\text{s}}}^{2}(t)\rightarrow 2v_{0}^{2}$, while it can be shown that for case (a) we have 
$\sigma_{xx}^{2}(t)\rightarrow 2v_{0}^2\tau_{R}t$, from which the active diffusion coefficient $D=v_{0}^{2}\tau_{R}$ is evident and $\sigma_{xv_{\text{s}}}^{2}(t)\rightarrow 
v_{0}^{2}\tau_{R}$. For the case (b) we have $\sigma_{xx}^{2}(t)\rightarrow 2v_{0}^2\tau_{R}\, (\gamma_{0}t)^{\beta}/\Gamma(\beta+1)\gamma_{0}$ and 
$\sigma_{xv_{\text{s}}}^{2}(t)\rightarrow v_{0}^{2}\tau_{R}(\gamma_{0}t)^{\beta-1}/\Gamma(\beta)$. $\gamma_{0}^{-1}$ is a timescale that characterizes the memory function and 
the relation $\sigma_{xv_{\text{s}}}^{2}(t)=(1/2)(d/dt)\sigma_{xx}^{2}(t)$ has been used.

Furthermore, in striking contrast with the zero-ranged memory function, which gives rise to positive correlations of the swimming velocity and to a smooth crossover between 
the ballistic superdiffusion and the normal diffusion, finite-ranged memory functions lead to anticorrelations of the swimming velocity in the intermediate-time regime. 
These anticorrelations are conspicuously revealed in the intermediate-time regime of $\sigma_{xx}^{2}(t)$, which are interpreted as a \emph{self-trapping effect}. This is 
discussed in detail in the following subsections.

The corresponding joint probability density of the active part of motion, $P_{\text{act}}\bigl(x,v_{\text{s}},t\vert x(0),v_{\text{s}}(0)\bigr)$, is 
given by the convolution of $G^{(2)}_{\text{act}}\left(x,v_{\text{s}},t\right)$ with the joint density induced by the deterministic part of Eqs. \eqref{glesols}, namely
\begin{multline}
 \int_{-\infty}^{\infty}dx^{\prime}\int_{-\infty}^{\infty}dv_{\text{s}}^{\prime}\, G^{(2)}_{\text{act}}\left(x-x^{\prime},v_{\text{s}}-v_{\text{s}}^{\prime},t\right)\\
 \times\delta\bigl(x^{\prime}-\langle x(t)\rangle\bigr)\delta\bigl( v_{\text{s}}^{\prime}-\langle v_{\text{s}}(t)\rangle \bigr).
\end{multline}
Using the characteristic function method~\cite{WangPRA1992}, one can easily show that $G^{(2)}_{\text{act}}\left(x,v_{\text{s}},t\right)$ satisfies the Fokker-Planck equation 
(see Appendix \ref{Appendix}),
\begin{multline}\label{ActiveFPE}
    \left(\frac{\partial}{\partial t}+v_{\text{s}}\frac{\partial}{\partial x}\right)G^{(2)}_{\text{act}}\left(x,v_{\text{s}},t\right)=\\
    \left(f(t)\frac{\partial}{\partial v_{\text{s}}}v_{\text{s}}+g(t)\frac{\partial^{2}}{\partial x\partial v_{\text{s}}}+h(t)\frac{\partial^{2}}{\partial v_{\text{s}}^{2}}\right)
    G^{(2)}_{\text{act}}\left(x,v_{\text{s}},t\right)
\end{multline}
where
\begin{subequations}\label{ActiveFPE-Coefficients}
\begin{align}
    f(t)&=\frac{\sigma^{2}_{v_{\text{s}}v_{\text{s}}}(t)}{\sigma^{2}_{xv_{\text{s}}}(t)},\\
    g(t)&=\frac{d}{dt}\sigma^{2}_{xv_{\text{s}}}(t),\\
    h(t)&=\frac{1}{2}\frac{d}{dt}\sigma^{2}_{v_{\text{s}}v_{\text{s}}}(t)+\frac{\left[\sigma^{2}_{v_{\text{s}}v_{\text{s}}}(t)\right]^{2}}{\sigma^{2}_{xv_{\text{s}}}(t)}.
\end{align}
\end{subequations}

In the following subsections we analyze the consequences of the present model by considering an instance of interest for each of the two asymptotic behaviors of $\gamma(t)$ 
considered in this paper. The first example considers a memory function that decays at least exponentially faster, while the second assumes the asymptotic behavior of a power 
law.

\subsection{Exponential memory kernel}

\begin{figure*}
 \includegraphics[width=\textwidth]{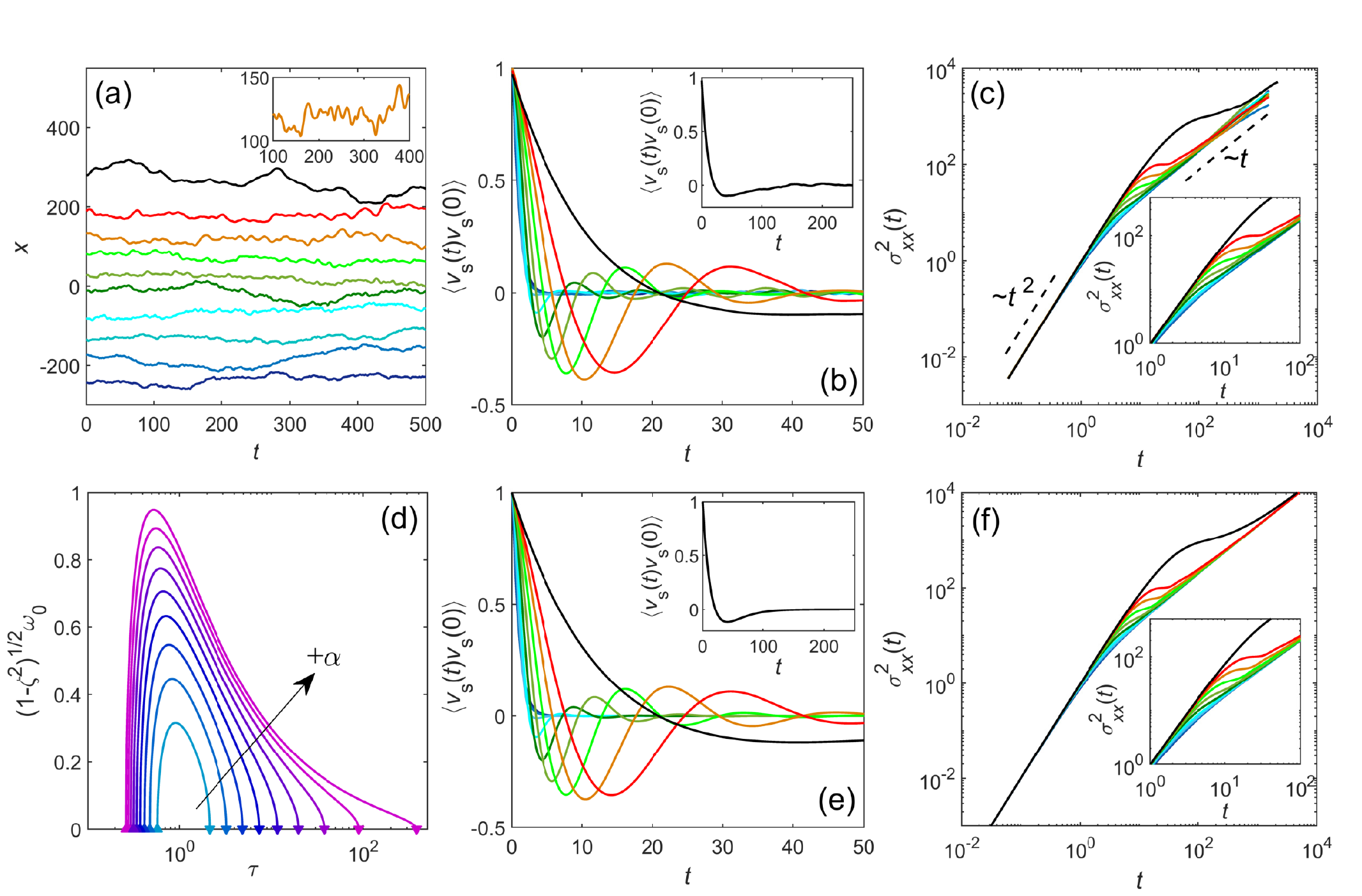}
 \caption{(a) Examples of trajectories $x(t)$ evolving according to the generalized Ornstein-Uhlenbeck model (\ref{gle}), with memory kernel given by (\ref{jeffreysgamma}), 
for $\alpha = 0.9$ and different values of the memory time $\tau$. From bottom to top: $\tau = 0.1, 0.2, 0.4, 0.8, 1.6, 3.2, 6.4, 12.8, 25.6, 400$. Inset: expanded view the 
active trajectory with $\tau = 12.8$. (b) Corresponding velocity autocorrelation function for different values of $\tau$, same color code as in (a). The values of $\tau$ 
increase from left to right. Inset: expanded view for $\tau = 400$. (c) Mean-squared displacements of the trajectories shown in (a). The values of $\tau$ increase from bottom 
to top. Inset: expanded view of the intermediate regime around $\tau_R$. (d) Dependence on the relaxation time $\tau$ of the frequency of the damped oscillations, 
which emerge only between $\tau_+$ ($\triangle$) and $\tau_-$ ($\triangledown$), for different values of $\alpha$ increasing from inner to outer curves: $\alpha = 0.1, 0.2, 
0.3, 0.4,0.5,0.6,0.7,0.8, 0.9$. (e) Velocity autocorrelation function and (f) mean-squared displacement, obtained from the analytical expression, for the same parameters 
plotted in (b) and (c), respectively.}
\label{FigPowerExp}
\end{figure*}

As a first example, we focus on a memory kernel consisting of a $\delta$ function plus an exponential decay with relaxation time $\tau$~\cite{FaEPJB2008},
\begin{equation}\label{jeffreysgamma}
    \gamma(t)  = 2 (1 - \alpha) \delta(t) + \frac{\alpha}{\tau}\exp\left( - \frac{|t|}{\tau}\right),
\end{equation}
where $0 < \alpha < 1$ is a dimensionless parameter that weighs the role of the exponential memory over the $\delta$ one. This kind of memory kernel describes the rheological 
response of several viscoelastic materials, such as intracellular fluids~\cite{WilhelmPRE2003}, polymer solutions~\cite{OchabSoftMatter2012}, wormlike 
micelles~\cite{SarmientoGomezJPCB2010}, and $\lambda$-phage DNA~\cite{GomezSolanoNJP2015}, where $\tau$ is the relaxation time of the elastic 
microstructure~\cite{PaulJPCM2018}. In the present work, it represents the retarded effects on the swimming velocity due to viscoelastic-like effects. More precisely, it 
considers two channels of persistence: the standard one, given by the $\delta$ function and considered in Ref. \cite{SzamelPRE2014}, that leads to exponentially decaying 
correlations of the swimming velocity, and the other one leads to long-lived correlations {exhibiting intermittently negative correlations in the 
intermediate-time regime}. For either $\alpha = 0$ or $\tau \rightarrow 0$, Eq.~(\ref{jeffreysgamma}) corresponds to the AOUM of Szamel~\cite{SzamelPRE2014}. 

In order to simulate trajectories evolving according to the generalized model presented in this paper, for $0 < \alpha  < 1$, we express Eq.~(\ref{glev}) in a Markovian form by introducing the additional variable
\begin{equation}\label{additionalvelocity}
    u(t)  = \frac{1}{\tau} \int_0^t ds \exp\left( -\frac{t - s}{\tau} \right)\left[ v_{\text{s}}(s) + \tau \phi_2(s) \right],
\end{equation}
where $\phi_2$ is a zero-mean Gaussian noise with autocorrelation
\begin{equation}\label{phi2}
   \langle \phi_2(t) \phi_2(s) \rangle = \frac{2v_0^2 \tau_R}{\alpha \tau^2} \delta(t-s).
\end{equation}
Then, Eq.~(\ref{glev}) can be written as
\begin{subequations}\label{jeffreysequations}
\begin{align}
\frac{d}{dt}v_{\text{s}}(t)&=-\frac{1-\alpha}{\tau_R} v_{\text{s}}(t) - \frac{\alpha}{\tau_R} u(t) + \phi_1(t) ,\label{jeff1}\\
\frac{d}{dt}u(t)&= -\frac{1}{\tau}[u(t) - v_{\text{s}}(t)] + \phi_2(t),\label{jeff2}
\end{align}
\end{subequations}
where $\phi_1(t)$ is a zero-mean Gaussian noise, which satisfies
\begin{equation}\label{phi1}
   \langle \phi_1(t) \phi_1(s) \rangle = \frac{2(1 - \alpha)v_0^2}{\tau_R} \delta(t-s).
\end{equation}

In the following, length scales are normalized by the \emph{persistence length} $v_0 \tau_R$, timescales by $\tau_R$, velocities by $v_0$, and translational diffusion 
coefficients by $v_0^2 \tau_R$. In Fig. \ref{FigPowerExp}(a) we plot some simulated trajectories for different values of the memory $\tau$ and constant $\alpha = 
0.9$. As $\tau$ increases, the shape of the trajectories change qualitatively, displaying three distinct kinds of behaviors.
To better appreciate such regimes for different values of $\tau$, we compute the corresponding velocity autocorrelation function $\langle v_{\text{s}}(t) v_{\text{s}}(0) 
\rangle$. {In accordance with our linear-response assumption, this is given by $v_{0}^{2}\Gamma^{\prime}(t)$ [see  Eq.~(\ref{ACFswimmingvel})], where $\Gamma^{\prime}(t)$ has 
been introduced in Eqs.~\eqref{glevsol}~and~\eqref{GammaPrime} and defined in Eq.~\eqref{LTGammaPrime}}. As shown in Fig. \ref{FigPowerExp}(b), for small values of $\tau$ 
($\tau=0.1$ and $\tau=0.2$) the velocity 
autocorrelation function $\langle v_{\text{s}}(t) v_{\text{s}}(0) \rangle$ exhibits a monotonic decay. Furthermore, damped oscillations of $\langle v_{\text{s}}(t) 
v_{\text{s}}(0) \rangle$ show up at larger $\tau$, thus manifesting the appearance of anticorrelations with a frequency that strongly depends on $\tau$, as observed for $0.4 
\le \tau \le 25.6$. {For instance, in the inset of Fig.~\ref{FigPowerExp}(a), such oscillations can be clearly observed along an active trajectory with $\tau = 
12.8$}. Moreover, the oscillations vanish at very large $\tau$, where $\langle v_{\text{s}}(t) v_{\text{s}}(0) \rangle$ exhibits a single global minimum, as shown 
in the inset of Fig. \ref{FigPowerExp}(b) for $\tau = 400$, where velocity anticorrelations occur. In Fig. \ref{FigPowerExp}(c) we show the resulting mean-squared 
displacements $\sigma^{2}_{xx}(t)$. For all values of the relaxation time $\tau$, a ballistic $\sigma^{2}_{xx}(t) \propto t^2$ and diffusive regime $\sigma^{2}_{xx}(t) \propto 
t$ is observed on timescales $t \ll \tau_R$ and $t \gg \tau_R$, respectively. This is in contrast to intermediate timescales (comparable to $\tau_R$), where a strong 
dependence on $\tau$ is found, see inset of Fig. \ref{FigPowerExp}(c).

Indeed, from Eqs. (\ref{jeffreysequations}), we can derive the following equation for the autocorrelation function:
\begin{multline}\label{harmonic_velcorr}
 \frac{d^2 \langle v_{\text{s}}(t)  v_{\text{s}}(0) \rangle}{d t^2} + \left( \frac{1}{\tau} + \frac{1-\alpha}{\tau_R} \right) \frac{d\langle v_{\text{s}}(t)  v_{\text{s}}(0) \rangle}{dt} \\
 + \frac{1}{\tau \tau_R} \langle v_{\text{s}}(t)  v_{\text{s}}(0) \rangle = 0,
\end{multline}
which is formally equivalent to the equation of motion of a damped harmonic oscillator with undamped angular frequency $\omega_0$ and damping ratio $\zeta$ given by
\begin{subequations}\label{harmonicparameters}
\begin{align}
    \omega_0 & = \frac{1}{\sqrt{\tau\tau_R}},\label{freq}\\
    \zeta = &\frac{1}{2}\sqrt{\tau \tau_R} \left(\frac{1}{\tau} + \frac{1-\alpha}{\tau_R}  \right),\label{rat}
\end{align}
\end{subequations}
respectively. Under the initial conditions $ \langle v_{\text{s}}(0)  v_{\text{s}}(0) \rangle = v_0^2$ and $\frac{d \langle v_{\text{s}}(t)  v_{\text{s}}(0) \rangle}{dt}|_{t = 0} = -\frac{1-\alpha}{\tau_R}v_0^2$, Eq. (\ref{harmonic_velcorr}) has three different kinds of solutions, which are determined by two particular values of the memory time $\tau$
\begin{subequations}\label{crittau}
\begin{align}
 \tau_{+} & = \frac{\tau_R}{(1+\sqrt{\alpha})^2},
  \\
 \tau_{-} & = \frac{\tau_R}{(1 - \sqrt{\alpha})^2} .
\end{align}
\end{subequations}
Note that $\tau_+ < \tau_R$, whereas $\tau_- > \tau_R$ for all values of $\alpha$. In particular, for the value $\alpha = 0.9$ considered here in most of our numerical 
results, $\tau_{+} = 0.26$ and $\tau_{-} = 379.74$. For $0 \le \tau < \tau_{+}$ or $ \tau_{-} < \tau$, the solution for  $\langle v_{\text{s}}(t)  v_{\text{s}}(0) \rangle$ is 
composed of two exponential decays, 
\begin{multline}\label{velcorr_exp}
    \langle v_{\text{s}}(t)  v_{\text{s}}(0) \rangle = \frac{v_0^2}{2\omega_0 \sqrt{\zeta^2 - 1}} \left[ A_{-} e^{ -\omega_0 \left(\zeta - \sqrt{\zeta^2-1}\right) t } \right.
 \\
 \left. + A_{+} e^{ -\omega_0 \left(\zeta + \sqrt{\zeta^2-1}\right) t} \right],
\end{multline}
where the amplitudes $A_{\pm}$ are given by
\begin{equation}\label{amplitudesoverdamped}
    A_{\pm} = \frac{1}{2}\sqrt{\left( \frac{1-\alpha}{\tau_R} + \frac{1}{\tau} \right)^2 - \frac{4}{\tau\tau_R}} \pm \frac{1}{2} \left( \frac{1-\alpha}{\tau_R} - \frac{1}{\tau}  \right).
\end{equation}
For $0 \le \tau < \tau_+$, Eq. (\ref{velcorr_exp}) represents a double-exponentially monotonic decay from $v_0^2$ to 0 of the velocity autocorrelation function.
This corresponds to the behavior shown in Fig.~\ref{FigPowerExp}(b) for $\tau = 0.1$ and 0.2, which are below $\tau_+$. On the other hand, $\tau_- < \tau$ yields a 
nonmonotonic dependence of $\langle v_{\text{s}}(t)  v_{\text{s}}(0) \rangle$ on $t$, with a single minimum around which anticorrelations $\langle v_{\text{s}}(t)  
v_{\text{s}}(0) \rangle < 0$ happen. This is illustrated in the inset of Fig.~\ref{FigPowerExp}(b) for $\tau = 400$, where  $\langle v_{\text{s}}(t)  v_{\text{s}}(0) \rangle < 
0$ for $t > 20.89$, while the minimum is located at $t = 40.72$.

At $\tau = \tau_{\pm}$, the velocity autocorrelation function takes the critical damping form
\begin{equation}\label{velcorr_crit}
   \langle v_{\text{s}}(t)  v_{\text{s}}(0) \rangle = v_0^2 e^{ -\frac{t}{\sqrt{\tau_{\pm} \tau_R}} } \left[ 1 + \left( \frac{1}{\sqrt{\tau_{\pm} \tau_R}} - \frac{1-\alpha}{\tau_R} \right) t \right],
\end{equation}
The two solutions (\ref{velcorr_crit}) separate the pure exponential solutions for $0 \le \tau < \tau_+$ and  $\tau_- < \tau$ from those
within the interval $\tau_{+} < \tau < \tau_{-}$. For the latter, the velocity autocorrelation function has the following damped-oscillatory form:
\begin{multline}\label{velcorrunderdamped}
   \langle v_{\text{s}}(t)  v_{\text{s}}(0) \rangle = v_0^2 \exp(-\zeta \omega_0 t) \left[ \cos\Bigl(\sqrt{1-\zeta^2}\omega_0 t\Bigr) \right. \\
    + \left. B \sin\Bigl(\sqrt{1-\zeta^2}\omega_0 t\Bigr) \right],
\end{multline}
where the amplitude
\begin{equation}\label{amplitudeunderdamped}
    B = \frac{1}{\sqrt{1 - \zeta^2}\omega_0}\left(\zeta \omega_0 - \frac{1-\alpha}{\tau_R} \right),
\end{equation}
and the frequency of the damped oscillations 
\begin{equation}\label{freqosc}
   \sqrt{1 - \zeta^2} \omega_0 = \sqrt{\frac{1}{\tau\tau_R} - \frac{1}{4} \left( \frac{1}{\tau} + \frac{1 - \alpha}{\tau_R}\right)^2},
\end{equation}
has a nonmonotonic dependence on $\tau$. This corresponds to the behavior observed for $\tau = 0.4, 0.8, 1.6, 3.2, 6.4, 12.8$ and 25.6 in Fig.~\ref{FigPowerExp}(b). In Fig. 
\ref{FigPowerExp}(d) we plot $\sqrt{1 - \zeta^2} \omega_0$ as a function of $\tau$ for different values of $\alpha$. While at small $\alpha$ the interval over which 
oscillatory solutions are possible is very narrow and the oscillation frequencies are low, it broadens and the corresponding frequencies are enhanced with increasing $\alpha$, 
i.e. when the exponential memory term in Eq. (\ref{jeffreysgamma}) becomes dominant. In Fig. \ref{FigPowerExp}(e) we show the velocity autocorrelation function obtained 
directly from the explicit expressions (\ref{harmonicparameters})--(\ref{freqosc}) for $\alpha = 0.9$ and the same values of $\tau$ as in \ref{FigPowerExp}(b), where 
excellent agreement with the numerical results is observed.

Using the previous expressions for $\langle v_{\text{s}}(t)  v_{\text{s}}(0) \rangle$, we can readily derive the corresponding ones for the mean-squared displacement. For $0 
\le \tau < \tau_{+}$ or $ \tau_{-} < \tau$, this reads
\begin{multline}\label{msdoverdamped}
    \sigma^{2}_{xx}(t)= 2v_0^2\tau_R \left[ t - (\tau_R - \alpha\tau) \right]
   \\
   +\frac{v_0^2}{\omega_0^2 \sqrt{\zeta^2-1}} \left[ C_{-}e^{-\omega_0\left(\zeta - \sqrt{\zeta^2 - 1}\right)t}  \right. \\ \left. + C_{+}e^{-\omega_0 \left(\zeta + \sqrt{\zeta^2 - 1}\right)t} \right] ,
\end{multline}
where
\begin{equation}\label{coeffmsdoverdamped}
    C_{\pm} = \pm \frac{-\zeta \pm \sqrt{\zeta^2 - 1} + \frac{1 - \alpha}{ \omega_0 \tau_R}}{2\zeta(\zeta \pm \sqrt{\zeta^2 - 1}) - 1}.
\end{equation}
At $\tau = \tau_{\pm}$, the expression for the mean-squared displacement is
\begin{multline}\label{msdcritdamped}
    \sigma^{2}_{xx}(t)= 2v_0^2\tau_R \Bigg\{ \left[ 1 \pm \frac{\sqrt{\alpha}}{1 \pm \sqrt{\alpha}} \exp \left( - \frac{1\pm \sqrt{\alpha}}{\tau_R} t\right) \right] t
   \\
   +\frac{1\pm 2\sqrt{\alpha}}{(1\pm \sqrt{\alpha})^2} \tau_R \left[ \exp \left( - \frac{1\pm \sqrt{\alpha}}{\tau_R} t\right) -1\right] \Bigg\},
\end{multline}
while for $\tau_{+} < \tau < \tau_{-}$, $\sigma^{2}_{xx}(t)$ can be expressed as 
\begin{multline}\label{msdunderdamped}
    \sigma^{2}_{xx}(t)= 2v_0^2\tau_R \Bigg\{ t - (\tau_R - \alpha\tau) 
   \\ + e^{-\zeta \omega_0 t} \Bigg[ \tau_R (\tau_R - \alpha\tau)  \cos(\sqrt{1-\zeta^2}\omega_0 t)
   \\  - \frac{2\zeta\sqrt{1 - \zeta^2} + (1-2\zeta^2)B}{\omega_0^2} \sin(\sqrt{1-\zeta^2}\omega_0 t) \Bigg] \Bigg\}.
\end{multline}
{Interestingly, mean-squared displacements which are similar to the critical damping (\ref{msdcritdamped}) and to the damped-oscillatory case (\ref{msdunderdamped}) have been 
observed for bacteria with run-reverse-flick swimming~\cite{TaktikosPlos2013}, for microorganisms with run-reverse locomotion~\cite{GrossmannNJP2016}, and for more general 
patterns of active motion \cite{Sevilla2019} or with a strong response to self-produced chemoattractants~\cite{TaktikosPRE2011}, respectively.}
In all cases, the previous expressions for the mean-squared displacement reduce to a ballistic regime $\sigma^{2}_{xx}(t) \rangle \approx v_0^2 t^2$ at short timescales, $t 
\ll \tau_R$. In contrast, at $t \gg \tau_R$ active diffusion $\sigma^{2}_{xx}(t) \approx 2D t$ is observed, where the active diffusion coefficient is $D = v_0^2 \tau_R$ for 
all values  of $\tau$, as shown in Figs. \ref{FigPowerExp}(c) for the numerical trajectories and in Fig. \ref{FigPowerExp}(f) for the analytical expressions. In the insets of 
Figs~\ref{FigPowerExp}(c) and \ref{FigPowerExp}(f), we show that the damped oscillations of $\langle v_{\text{s}}(t)  v_{\text{s}}(0) \rangle$ for $\tau_{+} < \tau < \tau_{-}$ 
translate into a shift of the short-time ballistic regime of $\sigma^{2}_{xx}(t)$ to timescales larger than $\tau_R$. For $\tau > \tau_-$, the ballistic behavior of 
$\sigma^{2}_{xx}(t)$ persists for timescales significantly larger than~$\tau_R$.

The effect of a nonzero thermal diffusion coefficient, $D_T = k_B T \mu> 0$, is to simply add an amount $2 D_T t$ to the mean-squared displacement of active motion, which 
results in a long-time active diffusion with coefficient $D_T + v_0^2 \tau_R$. Thus, such a diffusive behavior can be interpreted in terms of a nonequilibrium effective 
temperature $T_{\text{eff}} = T + \frac{v_0^2 \tau_R}{k_B \mu}$. Note that $T_{\text{eff}}$ increases quadratically with $v_0$ regardless of the value of the memory time 
$\tau$. This dependence is similar to that obtained from the conventional AOUM \cite{SzamelPRE2014} and also to that for active Brownian particles \cite{PalacciPRL2010}.

\subsection{Power-law memory kernel}

\begin{figure*}
 \includegraphics[width=\textwidth]{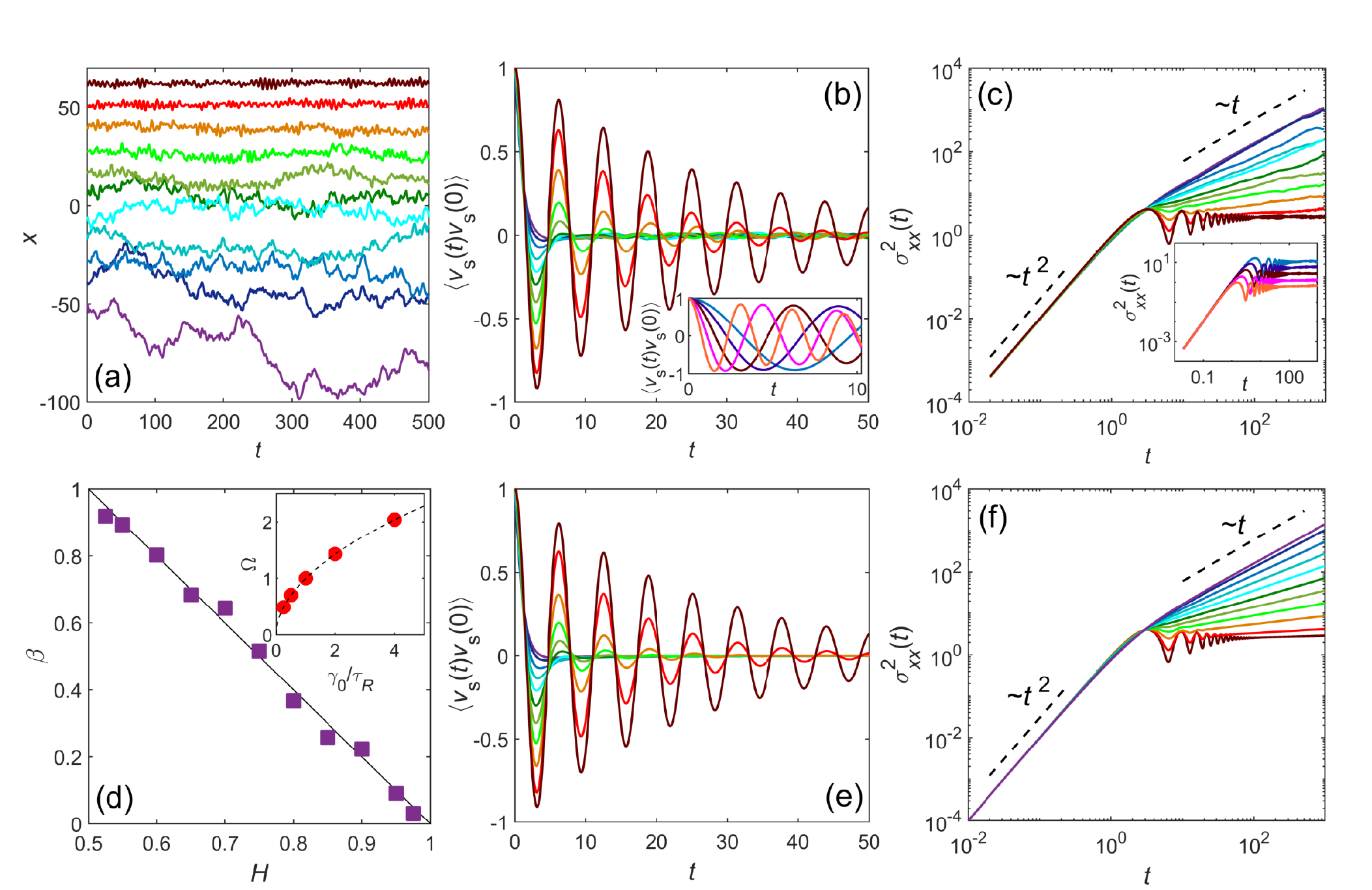}
 \caption{(a) Examples of trajectories $x(t)$ evolving according to the generalized Ornstein-Uhlenbeck model (\ref{gle}) with power-law memory kernel ~(\ref{powelawgamma}) and 
fractional Brownian noise (\ref{fBnvelocity}) in the absence of thermal fluctuations, for different values of the Hurst parameter increasing from bottom to top: $H = 0.525, 
0.55, 0.6, 0.65, 0.7, 0.75, 0.8, 0.85, 0.9, 0.95, 0.975$. (b) Velocity autocorrelation function and (c) corresponding mean-squared displacement for the different values of 
the Hurst parameter
 in (a), same color code. In (b) and (c), the values of
$H$ increase from inner to outer curves and from top to bottom, respectively. The insets in (b) and (c) corresponds to the 
velocity autocorrelation function and the mean-squared displacement for $H = 0.975$ and different values of $\gamma_0/\tau_R$; from left to right and bottom to top, 
respectively: $\gamma_0/\tau_R = 4, 2, 1, 0.5, 0.25$. (d) Exponent $\beta$ of the long-time behavior ($t \gg \tau_R$) of the mean-squared displacement as a function of $H$. 
The symbols ($\square$) mark the values obtained from the numerical solutions, whereas the solid line represents $2-2H$. Inset: Frequency of the velocity oscillations for $H 
= 0.975$ as a function of $\gamma_0/\tau_R$. The dashed line represents Eq. (\ref{freqPowerLaw}). (e) Velocity autocorrelation function and (f) mean-squared displacement, 
directly computed from the analytical expressions (\ref{velcorrpowerlaw}) and (\ref{msdpowerlaw}), respectively, for the same values of $H$ as those shown in 
(b) and (c).}
 \label{FigPowerLawH_DT0}
\end{figure*}

As a second example, we consider a power-law memory kernel \cite{RodriguezPhysA2015},
\begin{equation}\label{powelawgamma}
    \gamma(t) = \frac{\gamma_0 t^{2H-1}}{\Gamma(2H)}  \left[ \frac{2H - 1}{t} + 2 \delta(t)\right],
\end{equation}
where $\frac{1}{2} < H < 1$ guarantees the well-behaved time dependence of the quantities in Eqs.~\eqref{Gammas} and $\gamma_0 > 0$ a constant with units of time$^{1-2H}$. 
This kind of memory kernel describes several physical situations, such as the motion of granules within the cytoplasm \cite{TolicNorrelyke2004}, the micromechanical response 
of the cytoskeleton \cite{BallandPRE2006}, and rheological properties of soft biological tissues \cite{KobayashiPRE2017}. 
In this case, the corresponding stochastic term $\xi_{v_{\text{s}}}(t)$ in Eq.~(\ref{glev}) is a fractional Gaussian noise (characterized by the Hurst exponent $H$), with autocorrelation function
\begin{equation}\label{fBnvelocity}
    \langle \xi_{v_{\text{s}}}(t) \xi_{v_{\text{s}}}(s) \rangle = \frac{\gamma_0 v_0^2 |t - s|^{2H-1}}{\tau_R\Gamma(2H)}  \left[ \frac{2H - 1}{|t - s|} + 2 \delta(t-s)\right].
\end{equation}
{Note that this model corresponds to the conventional AOUM~\cite{SzamelPRE2014} if $H = \frac{1}{2}$}.
By integrating Eq. (\ref{glev}) over the time interval $[0,t]$, a straightforward calculation leads to the following expression for the velocity at time $t$:
\begin{equation}\label{velocityfBnmodel}
    v_{\text{s}}(t) = v_s(0) - \frac{\gamma_0}{\tau_R\Gamma(2H)}\int_0^t ds \frac{v_s(s)}{(t - s)^{1 - 2H}} + \chi(t),
\end{equation}
{where $\chi(t) = \int_0^t dt' \xi_{v_{\text{s}}}(t')$} is a fractional Brownian motion \cite{Qian2003}, which satisfies $\langle \chi(t) \rangle  = 0$ and 
\begin{equation}\label{fBmvelocity}
    \langle \chi(t) \chi(s) \rangle = \frac{\gamma_0 v_0^2}{2 \tau_R H \Gamma(2H)} \left( |t|^{2H} + |s|^{2H} - |t - s|^{2H} \right).
\end{equation}
We simulate particle trajectories evolving according to this generalized active Ornstein-Uhlenbeck model for different values of the parameters $H$ and $\gamma_0/\tau_R$. To 
this end, the integral on the right-hand side of Eq.~(\ref{velocityfBnmodel}) is evaluated using a modified Adams-Bashforth-Moulton algorithm \cite{DiethelmND2002}, 
whereas the fractional Brownian motion $\chi(t)$ is independently generated by means of the circulant embedding method of the covariance matrix \cite{DietrichSIAM1997}.

We first study the active motion of a free particle when no translational diffusion ($D_T = 0$) comes into play, i.e., $\frac{d}{dt}x(t) = v_{\text{s}}(t)$. The results for 
different values of $H$ are plotted in Figs. \ref{FigPowerLawH_DT0}(a)--(f), where length scales, timescales, velocities, and translational diffusion coefficients are 
normalized by $v_0 \tau_R$, $\tau_R$, $v_0$, and $v_0^2 \tau_R$, respectively. Some examples of simulated trajectories $x(t)$ for different values of $H$ and $\gamma_0 / 
\tau_R = 1$ are plotted in Fig. \ref{FigPowerLawH_DT0}(a). 
We find that with increasing $H$, the active trajectories develop a behavior ranging from quasidiffusion at $H$ slightly larger to $1/2$, to a strong self-trapping induced by 
persistent oscillations when $H$ is close to 1. Indeed, in Fig. \ref{FigPowerLawH_DT0}(b) we observe that the velocity autocorrelation function, $\langle 
v_{\text{s}}(t)v_{\text{s}}(0)\rangle$, exhibits a well-defined oscillatory behavior{, alternating between periods of positive correlations and negative 
correlations}, as $H$ increases. The frequency of the oscillations depends mainly on the parameter $\gamma_0 / \tau_R$, as confirmed in the inset of Fig. 
\ref{FigPowerLawH_DT0}(b) for $H = 0.975$. This can be understood from the fact that as $H$ approaches 1, the oscillations emerge from the competition between the long-range 
persistence of self-propulsion, described by the convolution in Eq.~(\ref{glev}), and the fractional Brownian noise $ \xi_{v_{\text{s}}}$. Since the intensity of the former is 
proportional to $\gamma_0 / \tau_R$, the quantity $(\tau_R/ \gamma_0)^{1/(2H)}$ sets the only characteristic timescale of the system, from which the frequency of the 
oscillations must be proportional to $(\gamma_0 / \tau_R)^{1/(2H)}$. Interestingly, the resulting mean-squared displacements display the typical ballistic regime 
$\sigma^{2}_{xx}(t) \propto t^2$ at short timescales $t \ll \tau_R$ for all $\frac{1}{2} < H < 1$, as shown in Fig. \ref{FigPowerLawH_DT0}(c). At larger timescales, the 
behavior of $\sigma^{2}_{xx}(t)$ strongly depends on $H$. For instance, for $H$ larger, but close to $\frac{1}{2}$, the mean square displacement exhibits approximately the 
long-time linear behavior expected for active Brownian motion: $\sigma^{2}_{xx}(t) \propto t$ for $t \gg \tau_R$. As $H$ increases, an intermediate oscillatory behavior at $t 
\gtrsim \tau_R$ shows up, where the amplitude of the oscillations of $\sigma^{2}_{xx}(t)$ eventually vanishes and leads to a subdiffusive growth at sufficiently large 
timescales, confirming the time dependence $\sigma^{2}_{xx}(t) \propto t^{\beta}$, with $\beta = 2 - 2H$ as shown in Fig. \ref{FigPowerLawH_DT0}(d). 
{ We point out that the previously described behavior is reminiscent of that of soft self-propelled particles with polar alignment in crowded glassy 
environments~\cite{HenkesPRE2011} and active particles in disordered heterogeneous media~\cite{ChepizhkoPRL2013,MorinPRE2017}. In such cases,
interparticle and alignment interactions induce long-range temporal correlations in the swimming velocity, which in turn lead to local trapping of the particles, thereby 
exhibiting transient oscillations  followed by long-time subdiffusion.}

An analytical expression for the velocity autocorrelation function can be derived from the general solution of Eq. (\ref{glev}), given by Eqs.~ (\ref{glevsol}), 
(\ref{GammaPrime}), and (\ref{LTGammaPrime}). In this case, the Laplace transform of the power-law memory kernel (\ref{powelawgamma}) is explicitly given by 
$\tilde{\gamma}(\epsilon) = \gamma_0 \epsilon^{1 - 2H}$.
Then a straightforward calculation leads to
\begin{equation}\label{velcorrpowerlaw}
    \langle v_{\text{s}}(t) v_{\text{s}}(0) \rangle = v_0^2 E_{2H,1}\left( - \frac{\gamma_0 t^{2H}}{\tau_R }\right),
\end{equation}
where $E_{\mu,\nu}(z)$ is the two-parameter Mittag-Leffler function, defined by the series expansion 
\begin{equation}\label{MLfunct}
    E_{\mu,\nu}(z) = \sum_{k = 0}^{\infty} \frac{z^k}{\Gamma(\mu k + \nu)},
\end{equation}
with $\mu>0$ and $\nu>0$.
In Fig. \ref{FigPowerLawH_DT0}(e) we demonstrate that the velocity autocorrelation curves computed from Eq. (\ref{MLfunct}) reproduce very well the numerical results of Fig. 
\ref{FigPowerLawH_DT0}(b) for all the values $H$. In particular, we note that $E_{1,1}(z) = \exp(z)$, while $E_{2,1}(-z^2) = \cos(z)$. Therefore, as $H \rightarrow 
\frac{1}{2}$, the velocity autocorrelation tends to the conventional Ornstein-Uhlenbeck model, $\langle v_{\text{s}}(t) v_{\text{s}}(0) \rangle = v_0^2 \exp(-\gamma_0 t / 
\tau_R)$, with relaxation time $\tau_R / \gamma_0$, where $\gamma_0$ is a dimensionless parameter. On the other hand, as $H$ approaches 1, $\langle v_{\text{s}}(t) 
v_{\text{s}}(0) \rangle$ develops a slow-decaying oscillatory behavior with frequency 
\begin{equation}\label{freqPowerLaw}
\Omega = \left(\frac{\gamma_0}{\tau_R} \right)^{\frac{1}{2H}},
\end{equation}
in agreement with the frequencies computed numerically, as verified in the inset of Fig. \ref{FigPowerLawH_DT0}(b) for $H = 0.975$.

\begin{figure*}
 \includegraphics[width=\textwidth]{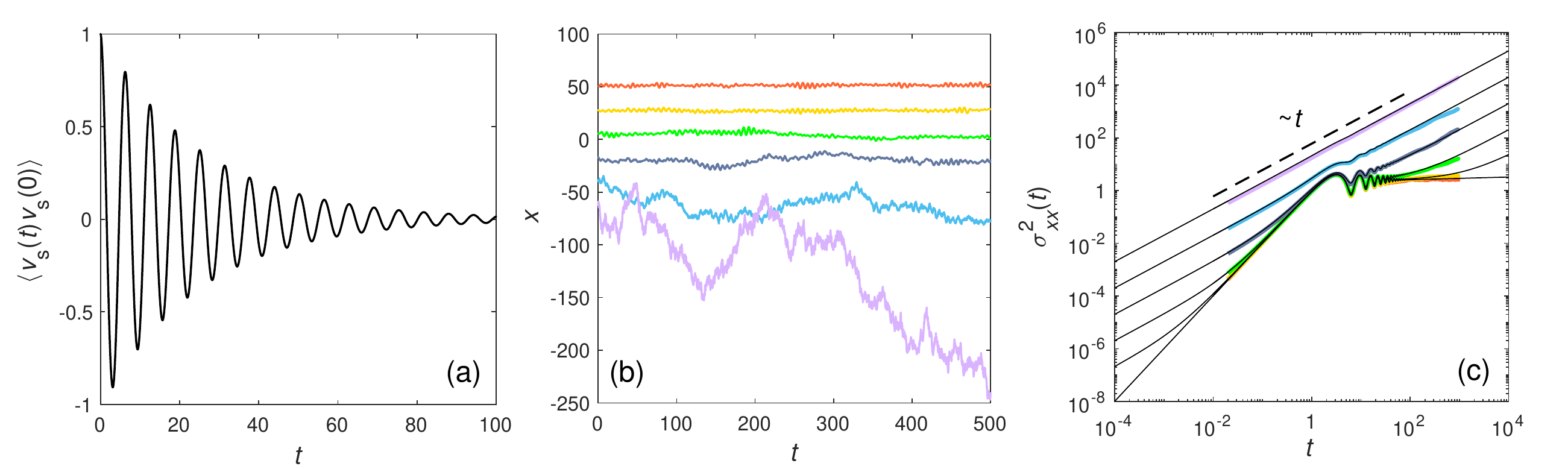}
 \caption{(a) Velocity autocorrelation function for the active Ornstein-Uhlenbeck model with the power-law memory kernel \eqref{powelawgamma} and fractional Brownian noise at 
$H = 0.975$. (b) Resulting active trajectories for different translational diffusion coefficients, from top to bottom: $D_T = 0, 10^{-3}, 10^{-2}, 10^{-1}, 1, 10$. (c) 
Corresponding mean-squared displacements. Same color code as in Fig. \ref{FigPowerLawDT}(b). The values of $D_T$
increase from bottom to top. The solid lines represent Eq. (\ref{msdpowerlawDT}).}
 \label{FigPowerLawDT}
\end{figure*}

In a similar manner, using the general solution for the particle position given by Eq. (\ref{glexsol}), we obtain the following expression for the mean-squared displacement:
\begin{equation}\label{msdpowerlaw}
    \sigma^{2}_{xx}(t) = 2v_0^2 t^2 E_{2H,3}\left( -\frac{\gamma_0 t^{2H}}{\tau_R} \right).
\end{equation}
Once again, Eq. (\ref{msdpowerlaw}) agrees very well with our numerical results shown in Fig. \ref{FigPowerLawH_DT0}(c) for all $H$, see Fig. \ref{FigPowerLawH_DT0}(f). For 
instance, for $t \ll (\tau_R / \gamma_0)^{\frac{1}{2H}}$, $E_{2H,3}(-z^{2H}) \approx 1/\Gamma(3) = 1/2$ regardless of $H$, and thus Eq. \eqref{msdpowerlaw} reduces to the 
short-time ballistic regime, $\sigma^{2}_{xx}(t) \approx v_0^2 t^2$.
{It should be noted that the oscillations of $\sigma^{2}_{xx}(t)$ for $t > \tau_R$ with increasing $H$ can only be captured when taking into account the full solution of the 
velocity autocorrelation function given in terms of the Mittag-Leffler functions, see Eq. (\ref{velcorrpowerlaw}). The oscillatory behavior of $\sigma^{2}_{xx}(t)$ is 
smeared out by any asymptotic power-law approximation of $\langle v_{\text{s}}(t) v_{\text{s}}(0) \rangle$, as those considered in Ref. \cite{WangPRA1992}.} Furthermore, 
taking into account the asymptotic behavior of the general Mittag-Leffler function $E_{\mu,\nu}(-z) \approx z^{-1}/\Gamma(\nu - \mu)$ for $z \rightarrow \infty$, the 
long-time behavior $\bigl[t \gg (\tau_R / \gamma_0)^{\frac{1}{2H}}\bigr]$ of the mean-squared displacement is \cite{Pottier2003,SevillaChapter2018}
\begin{equation}\label{msdpowerlawlongtime}
    \sigma^{2}_{xx}(t) \approx \frac{2v_0^2 \tau_R}{\gamma_0 \Gamma(3 - 2H)}  t^{2-2H},
\end{equation}
thereby reproducing the exponent $\beta$ of the active subdiffusive regime we find numerically, see Fig. \ref{FigPowerLawH_DT0}(d). In particular, from Eq. 
(\ref{msdpowerlawlongtime}) we recover the long-time dependence $\sigma^{2}_{xx}(t) \approx 2v_0^2 (\tau_R / \gamma_0) t$ as $H \rightarrow \frac{1}{2}$, while the active 
motion is subdiffusive with exponent $2 - 2H$ for $H > \frac{1}{2}$. Total spatial self-trapping occurs for complete persistence, i.e., for $H=1$, for which the mean-squared 
displacement saturates to the value $\sigma^{2}_{xx}(t \rightarrow \infty) =  2v_0^2 \tau_R / \gamma_0$.

In order to better illustrate the effect of thermal fluctuations on the active trajectories, we focus on a large value of the Hurst parameter ($H=0.975$), for which the 
velocity autocorrelation function exhibits a pronounced oscillatory behavior, see Fig.~\ref{FigPowerLawDT}(a). The overall effect is that the presence of a nonzero $D_T > 0$ 
destroys the long-time subdiffusive behavior, thus leading to trajectories with a large dispersion compared to the diffusion-free case, as shown in 
Fig.~\ref{FigPowerLawDT}(b). In fact, in the presence of translational thermal noise, the mean squared displacement is supplemented by a diffusive term $2D_T t$,
\begin{equation}\label{msdpowerlawDT}
    \sigma^{2}_{xx}(t) = 2v_0^2 t^2 E_{2H,3}\left( -\frac{\gamma_0 t^{2H}}{\tau_R} \right) + 2D_T t.
\end{equation}
Thus, depending on the value of $D_T$ and the timescale $t$, different regimes are observed. Indeed, in Fig.~\ref{FigPowerLawDT}(c), we observe that at short timescales, the 
mean-squared displacement has a diffusive part (diffusion coefficient equal to $D_T$), because the ballistic motion is negligible with respect to thermal diffusion. 
Furthermore, at sufficiently low $D_T$, typically $D_T \lesssim v_0^2 \tau_R$, and intermediate timescales (comparable to $\tau_R$), the oscillatory regime is still observed. 
On the other hand, for a sufficiently large thermal diffusion coefficient ($D_T \gtrsim v_0^2 \tau_R$), diffusion dominates completely the particle motion over all timescales, 
thereby hindering the memory-induced oscillations. For all values of $D_T$, the long-time diffusive behavior occurs, i.e., $\sigma^{2}_{xx}(t) \approx 2 D_T t$ for $t \gg 
\tau_R$, due to the dominance of thermal diffusion over the subdiffusive growth $t^{2-2H}$ in the mean-squared displacement. In all cases, Eq. (\ref{msdpowerlawDT}) perfectly 
describes our numerical results over all timescales and for all values of $H$, see solid lines in Fig.~\ref{FigPowerLawDT}(c).

We want to point out that in the case of the long-ranged memory kernel considered here, unlike the case of the finite-ranged one given in Eq. (\ref{jeffreysgamma}), the interpretation of the long-time limit of \eqref{msdpowerlawDT} in terms of an effective temperature is less clear. 
In fact, if $D_T = 0$, then an effective temperature cannot be defined in a straightforward manner, mainly due to the long-ranged (anti-)correlations of the swimming 
velocity that leads to a self-trapping effect and therefore to the subdiffusive behavior of the mean-squared displacement (\ref{msdpowerlawlongtime}). On the other hand, for 
$D_T > 0$ and in the long-time regime, the thermal fluctuations overcome the long-ranged correlations of the swimming velocity induced by the memory function. Therefore, the 
effective temperature of the resulting diffusive process exactly equals the temperature $T$ of the bath regardless of $v_0$, see Eq. (\ref{msdpowerlawDT}).

\section{\label{SectIV} Summary and final remarks}

In this work, we have investigated a generalization of the so-called active Ornstein-Uhlenbeck model for the motion of self-propelled particles subject to both thermal and 
nonequilibrium active fluctuations. The model considered here is based on the generalized Langevin equation \eqref{glev} for the swimming velocity and incorporates different 
channels of persistence of the particle swimming velocity by means of a memory function and additive colored noise. 
We have explicitly obtained the joint
probability density of the particle position and its swimming velocity for the complete process. We have also shown that such a probability density can be split into a thermally diffusive component and an active one. The latter satisfies the Fokker-Planck equation \eqref{ActiveFPE}, which explicitly involves the time-dependent elements of the active covariance matrix.

We have obtained numerical and analytical results for the velocity autocorrelation function and the mean-squared displacement for
two specific  memory functions that arise in many natural systems: a finite-ranged exponential decay
and a long-ranged power law. In both cases, damped-oscillatory behavior, that alternates between positive and negative correlations, of the swimming velocity emerges for 
certain values of the relevant parameters. The oscillations are damped in the case of the exponential decay, which leads to the emergence of an active diffusion coefficient 
and allows the definition of a nonequilibrium effective temperature. In contrast, oscillations are long lived for the power-law memory, and, remarkably, long-time 
subdiffusion is observed. This provides a simple example of free self-propelled motion where the concept of nonequilibrium effective temperature can not be trivially applied.

Although the effects of exponential memory have already been explicitly considered on the rotational motion of active Brownian particles \cite{NarinderPRL2018,GhoshJChemPhys2015,PeruaniPRL2007,HuJSTAT2017}, to our knowledge this is the first time that a general formulation encompassing long-lived correlations in the swimming speed has been studied. Our approach has allowed us to uncover numerous patterns of active motion which are absent in the conventional AOUM. Therefore, we expect that our results will be relevant for the understanding and modeling of intricate active systems, whose underlying dynamics, caused either by internal or external mechanisms, give rise to strong memory effects.
{ In fact, our single-particle model is able to qualitatively capture a variety of behaviors observed in numerous active systems where long-range memory in the swimming 
velocity emerges either from self- or interparticle interactions.
Similar effects are also expected to happen for deformable, asymmetric, or chiral self-propelled particles swimming in non-Newtonian fluid environments. Under such conditions, 
the local rheological properties of the medium, coupled to the response of the particle, can result in strongly correlated fluctuations of the propulsion velocity.}
A further step will be to investigate the effect of confining potentials and external flows, as they introduce additional timescales and correlations that could significantly 
modify the persistence of the active motion. 

\acknowledgments
F.J.S. kindly acknowledges support from DGAPA,  UNAM-PAPIIT-IN114717.

\appendix
\section{\label{Appendix}Derivation of the Active Fokker-Planck Equation}
We briefly derive the Fokker-Planck equation \eqref{ActiveFPE} for the bivariate probability density, $G^{2}_{\text{act}}(x,v_{\text{s}},t)$, that corresponds to the active 
part of motion. The starting point is the characteristic function $\hat{G}_{\text{act}}^{(2)}(k,q,t)$ of active motion, given by Eq. \eqref{ActG}. After applying the advective 
derivative in Fourier space, $\frac{\partial}{\partial t}-k\frac{\partial}{\partial q}$, to the expression \eqref{ActG} we have that
\begin{multline}\label{ActiveFPE-Fourier}
    \Biggl(\frac{\partial}{\partial t}-k\frac{\partial}{\partial q}\Biggr)\hat{G}_{\text{act}}^{(2)}(k,q,t)=\\
    -qk\Biggl(\frac{d}{dt}\sigma_{v_{\text{s}}v_{\text{s}}}^{2}(t)-\sigma_{xv_{\text{s}}}^{2}(t)\Biggl)\hat{G}_{\text{act}}^{(2)}(k,q,t)\\
    -q^{2}\Biggl(\frac{1}{2}\frac{d}{dt}\sigma_{v_{\text{s}}v_{\text{s}}}^{2}(t)\Biggr)\hat{G}_{\text{act}}^{(2)}(k,q,t),
\end{multline}
where $\sigma_{v_{\text{s}}v_{\text{s}}}^{2}(t)$, $\sigma_{xv_{\text{s}}}^{2}(t)$ and $\sigma_{xx}^{2}(t)$ are the elements of the active covariance matrix 
$\Sigma_{\text{act}}$, and we have used that $\sigma_{xv_{\text{s}}}^{2}(t)=\frac{1}{2}\frac{d}{dt}\sigma_{xx}^{2}(t)$, which makes the proportional terms to $k^{2}$ cancel 
each other. By noticing that 
\begin{multline}
    qk\, \hat{G}_{\text{act}}^{(2)}(k,q,t)=-\frac{1}{\sigma_{xv_{\text{s}}}^{2}(t)}
    q\frac{\partial}{\partial q}\hat{G}_{\text{act}}^{(2)}(k,q,t)\\
    -q^{2}\, \frac{\sigma_{v_{\text{s}}v_{\text{s}}}^{2}(t)}{\sigma_{xv_{\text{s}}}^{2}(t)}\hat{G}_{\text{act}}^{(2)}(k,q,t),
\end{multline}
(as can be checked straightforwardly by direct substitution), we have that Eq. \eqref{ActiveFPE-Fourier} can be rewritten as
\begin{multline}\label{ActiveFPE-Fourier2}
    \left(\frac{\partial}{\partial t}+k\frac{\partial}{\partial q}\right)\hat{G}^{(2)}_{\text{act}}\left(k,q,t\right)=-
    \Biggl(f(t)q\frac{\partial}{\partial q}\\
    +g(t)\, qk+h(t)q^{2}\Biggr)
    \hat{G}^{(2)}_{\text{act}}\left(k,q,t\right),
\end{multline}
whose inverse Fourier transform directly leads to the Fokker-Planck equation \eqref{ActiveFPE}, with $f(t)$, $g(t)$, and $h(t)$ as given in Eqs. 
\eqref{ActiveFPE-Coefficients}.

\providecommand{\noopsort}[1]{}\providecommand{\singleletter}[1]{#1}%

\end{document}